# An empirical model for the vegetation screening effect in the remote sensing monitoring

V. F. KRAPIVIN[1], C. A. VAROTSOS[2] and S. V. MARECHEK[1]

[1] Kotelnikov Institute of Radioengineering and Electronics, Russian Academy of Sciences, Moscow, Russia.
[2] Department of Environmental Physics and Meteorology, University of Athens, Athens, Greece.

**Abstract.** A layer of vegetation over the soil surface absorbs microwave radiation emitted from the soil and some objects located under the vegetation canopy. This paper investigates the microwave propagation process within the vegetation layer and evaluate the attenuation coefficients. For this purpose, laboratory experimental test bench is synthesized. It is based on the special rectangular camera-waveguide with transmission and receiving wide band horn antennas and allows the study of the attenuation coefficient dependence on the frequency of 0.8-10 GHz range and water content in the different vegetation elements. Evaluations of attenuation coefficients are implemented over a wide frequency range for deciduous and coniferous forests with separation consideration of the branches and trunks as the canopy fragments.

## 1. Introduction

Microwave radiometric tools of remote sensing are used to diagnose the soil-plant formations, snow and ice covers, oceanic surfaces, atmosphere and different natural-technogenic objects. To solve inverse tasks of the radiometry it is necessary to estimate the parameters of object that is diagnosed. In this case, knowledge of the attenuation characteristics of electromagnetic waves under transition across the vegetation covers is important for the implementation of the inverse tasks and for the reliable radio communication. Attenuation of electromagnetic waves of microwave range in the vegetation layer is a defining factor under the study of radiation and dispersion of radio waves in the vegetation layers. Moreover, data about the depending attenuation parameters on the frequency, vegetation biomass and water content, as well as on the structural characteristics of the vegetation covers give the basis for the reconstruction of the environment where radio waves are propagated.

General approach to the description of propagation of electromagnetic waves through vegetation layers is mainly based on the appropriate model creation that usually reflects heterogeneity of geometrical form and proportions of the vegetation layer elements. There is no doubt that propagation model of electromagnetic waves in vegetation canopy demands the knowledge of the configuration of branches, trunks and leaves. Theoretical models usually assume that these elements are mostly parallel to each other but are somewhat random in their lateral positions or needle cluster is oriented along the vertical axis of local coordinate system [1-4]. In general, the vegetation model requires as input information about the geometry, biomass and water content of canopy. A canopy geometric characteristics have spatial multiformity what complicates the model synthesis.

For the microwave range, sizes of trunks, leafs and branches are compared with wavelengths what determines basic complications when the propagation model is synthesized. In this case, it is necessary to take into consideration of a wide variety of geometries including single trees, lines of trees, and dense woodland, tree spacing, heights and leaf dimensions [5-7]. The problem multi-face complicates its solution and therefore approximate models are used when it is supposed that discrete scatterers have a random distribution in the space [8].

Usually, scalar radiative transfer equation is formed solution of which is realized under typical assumptions about a canopy structure. Similar models need the evaluation of their parameters what can be made by experimental measurements. For example, a model for the vegetation attenuation of microwaves has been proposed [9] taking into account the total cross section for leaves and branches and basing on Rayleigh scattering and geometric optics laws depending on the frequencies. There are several approaches to the study of the propagation of radio waves in the vegetation including theoretical and empirical tools [10]. Experimental results of the measurements of radio signal attenuation under its propagation through vegetation at three frequencies (1.3, 2.0 and 11.6 GHz) considering different geometries and tree species are given in [11]. These measurements were used to obtain the nonzero gradient model that predicts the excess attenuation due to vegetation. The results obtained from the experimental investigation of a forest transmission coefficient at 6.8-18.7 GHz frequency range as the function of the volume of forest trunks are presented in [12]. A model based on the radiative transfer theory and the matrix doubling algorithm has been presented in [13]. The forest medium is represented as morphological structure from canopy, trunks and soil where canopy consists from leaves, needles, twigs and branched positioned at various height and described by geometrical shares such as discs, ellipses or circles, cylinders, spheres and cones. Usually a representation of canopy structure is based on continuous statistical distribution of these geometric shares [14, 15].

Theoretical models describing the attenuation effect of the microwave propagation in vegetation canopy need the verification and validation what usually realized by means of synchronous experimental measurements of basic model parameters. In considered case, the most important outcome is search empirical model that could diminish variety of theoretical models. Existing experimental data about the attenuation radio waves in the vegetation layers were received for restricted ranges [16-18]. Therefore, in this paper the laboratory rectangular camera-waveguide is constructed to measure the continuous attenuation spectrums for the fragments of vegetation covers in the range of 0.8-10 GHz under controlled biometric characteristics. These measurements allow the empirical model synthesis.

## 2. Measuring system for retrieving attenuation of microwaves in the vegetation

Structure and principal scheme of the measuring system for retrieving attenuation of microwaves in vegetation (MSRAMV) is represented in Figure 1 and Table 1. Measuring process is taken priority of the assessment of basic electric characteristics for the MSRAMV microwave tract including two coaxial cables, two wide band horn antennas, and rectangular camera-waveguide. Figure 2 characterizes the frequency dependence of the attenuation coefficient for the null camera. It is follows that attenuation coefficient does not exceed 5 dB in the 0.8-2.15 GHz range.

Measuring process consists from the following stages:

• Choice of vegetation type and preparation of the plant fragments forming the separate completes of trunks and branches with or without leafs. These completes are weighed and their water content is evaluated. After that a complete of vegetation fragments is placed to the MSRAMV camera.

• Calibrated value of microwave power is generated on the output of P2-102, after that it enters through VSWR (Voltage Standing Wave Ratio) bridge spectrum analyzer to the antenna No.1 input where TEM wave is transformed to the $H_{10}$ wave that is propagated in the MSRAMV camera.

- The $H_{10}$ wave is transmitted through studied vegetation samples and is partly absorbed within them. Relaxed $H_{10}$ wave and generated waves of the highest order are transformed by the antenna No.2 to the TEM wave type.
- Registered data are analyzed with the using the algorithms and models mentioned in Table 1.

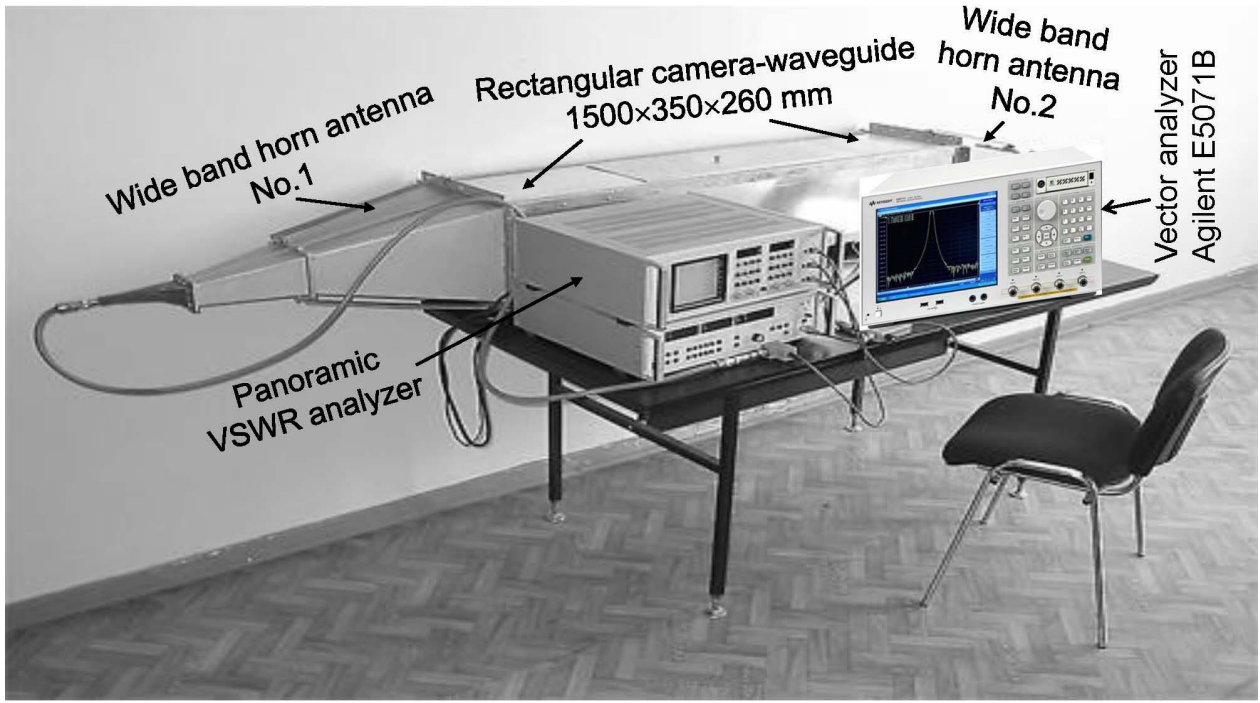

**Fig. 1.** Structure of measuring system for the attenuation evaluation of electromagnetic waves in the 0.8-10.0 GHz band. Functions of the MSRAMV are described in Table 1.

**Table 1.** Description of the MSRAMV functional components.

| MSRAMV component | Component functional destination |
|---|---|
| Rectangular camera-waveguide. | It means for the placing vegetation fragments including trunks, leaves, branches, etc. Camera has a length of 1500 mm, height of 350 mm, and width of 260 mm. Camera was produced from stainless steel of 1mm in the thickness. |
| Wide band horn antenna No. 1. | This is transmitting antenna that transforms a transverse electromagnetic (TEM) wave to the $H_{10}$ wave under small excitation level of the waves of the highest orders. |
| Wide band horn antenna No.2. | This is receiving antenna that realizes inverse S-transformation of $H_{10}$ wave and waves of the highest level to the TEM waves. |
| Panoramic VSWR analyzer P2-102. | It measures attenuation and stationary wave coefficient by voltage and generates the tilted frequency of the 10-2150 MGz range. |
| Vector analyzer Aligent E507 1B. | It measures the attenuation of electromagnetic waves when frequency is changed continuously in the 0.0003-10.0 GHz range. |
| Algorithms and models. | Algorithm for the polynomial approximation of experimental data. Algorithm for the inverse task solution of microwave radiometry. Two-level model of vegetation cover [19]. |

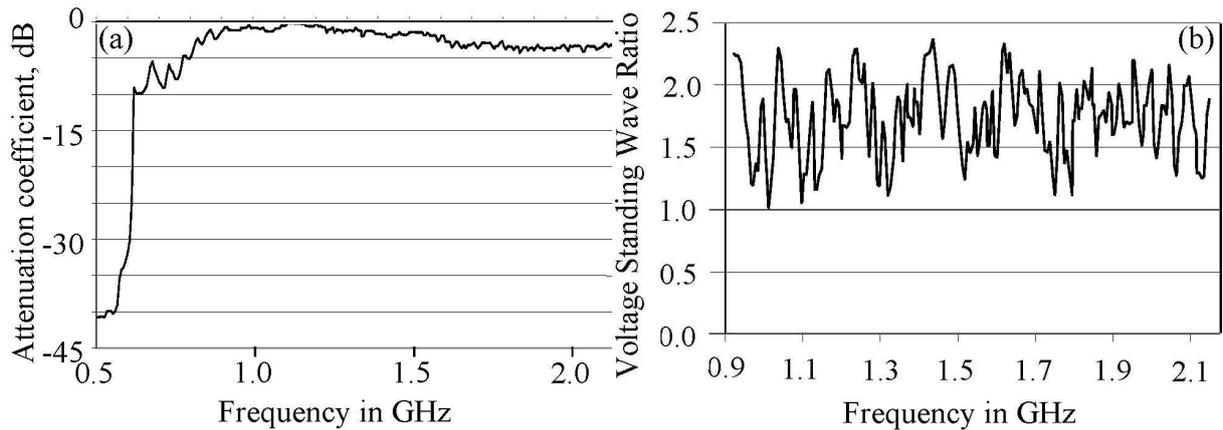

**Fig. 2.** Basic SHF-tract characteristics. Frequency dependence of the attenuation coefficient for SHF-tract (a) and of the VSWR at the SHF-tract input (b).

The MSRAMV calibration was realized using the camera filling with the etalon absorber such as foam-rubber having known dielectric characteristics and with introduction to it's the conductive carbonic material. A level of possible SHF-power transmission from antenna No.1 to antenna No.2 bypassing the camera is assessed by means of the camera filling with absolutely black solid such as material with large optical thickness. In this case, attenuation coefficient exceeded 55 dB. The MSRAMV precision is evaluated as the discrepancy between these measurements and theoretical calculations. Main contribution to the MSRAMV error gives the waves of the highest order subordinated on the camera imperfections. Maximal error did not exceed of 10%. At the same time the VSWR was measured for the null and filled camera the results of which were identical what means a minimal error due to the SHF tract non-coordination.

### 3. Experimental results

The MSRAMV is used to study spectral dependency of the attenuation characteristics for propagation of electromagnetic waves in the vegetation layer on its biometric parameters in the 0.8-10.0 GHz band. Measurement procedure includes the formation of completes of the fresh-cutting off branches and other vegetation elements a humidity of which corresponds to their natural state. Distribution of the vegetation elements within the camera corresponds to the state of forest canopy for different trees and their volumetric part does not exceed 0.02-0.03 what is specific for the top layer of natural forest. Gravimetric moisture of fresh-cutting off vegetation elements is evaluated after drying or after controlling drying. Some experimental results are given in Figures 3-6.

Fig. 3 demonstrates an example of attenuation coefficient dependence on the frequency for pine branches of different gravimetric moisture. Dry pine branches are characterized by the attenuation coefficient changes depending on the gravimetric moisture by 0.69 dB\kg\m$^2$ in low-wave band and by 1.74 dB\kg\m$^2$ in high-wave band. For fir-tree crone attenuation coefficient is less by 0.9%. Averaged attenuation coefficient $b$ (dB/kg/m$^2$) dependence on the frequency $f$ (GHz) can be approximated by the following formula: $b=cf^\beta$ where coefficients $c$ and $\beta$ are changed from 1.26 to 5.92 and from 0.72 to 2.04, respectively, when gravimetric moisture of the branches is changed from 17.1 to 62.7%.

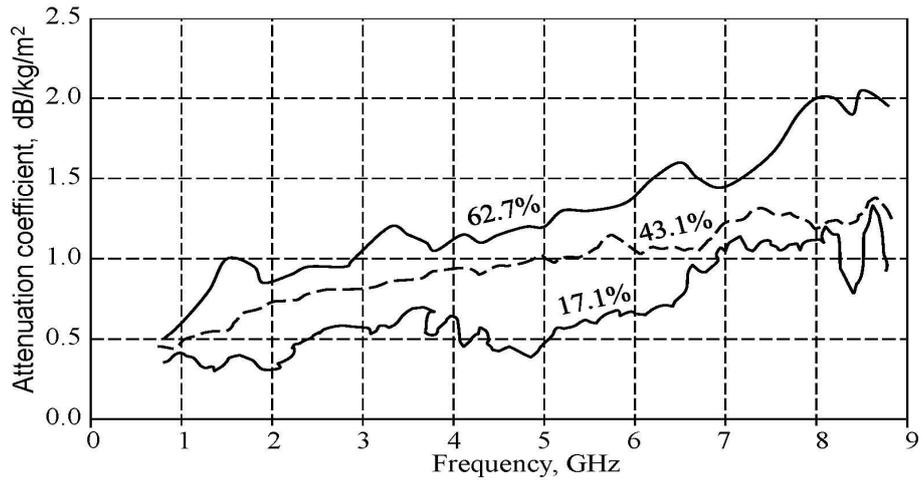

**Fig.3.** Attenuation coefficient dependency on the frequency and water content (it is shown on the curves in the percents) in the pine branches.

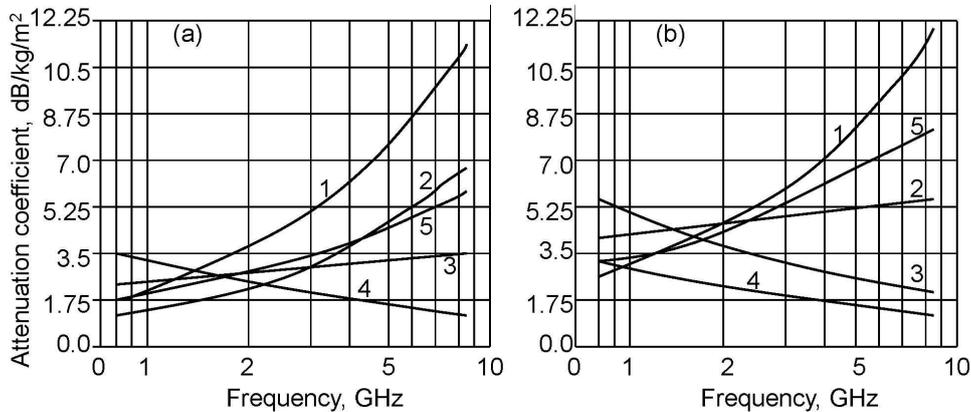

**Fig. 4.** Attenuation coefficient dependence on the aspen (a) and maple (b) branches for different their fragments: 1 – branches with the leaves of up to 5 mm diameter (gravimetric moisture is 47%); 2 – branches without leaves of up to 5 mm diameter (46%); 3- branches of 5-20 mm diameter (45%); 4 – branches of 20-50 mm diameter (45%); 5 – natural canopy (46%).

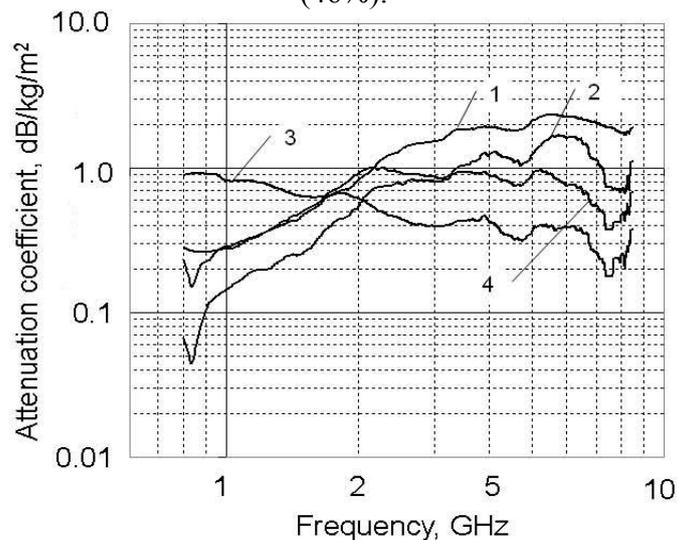

**Fig. 5.** Attenuation coefficient dependence on the frequency for different aspen branches. Notation: 1 – branches with leaves of 5 mm diameter; 2 – branches without leaves of 5 mm diameter; 3 – branches of 20-50 mm diameter; 4 – branches of 5-20 mm diameter.

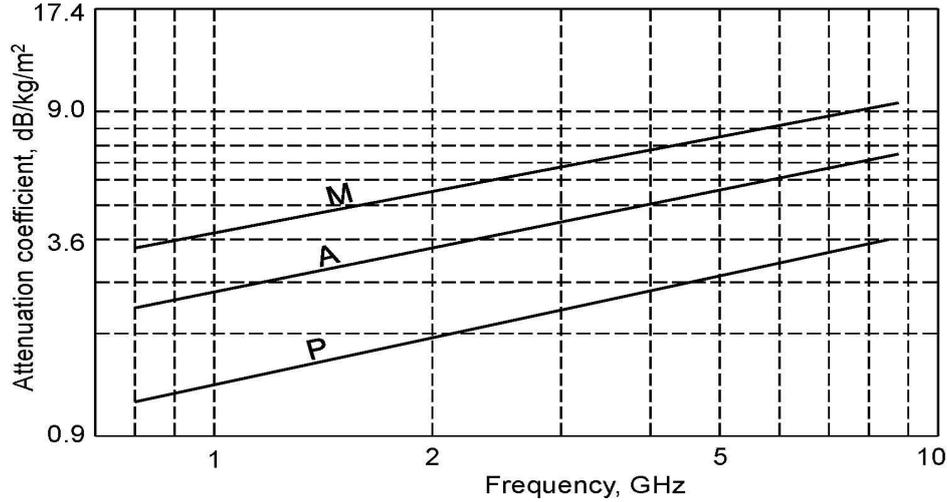

**Fig. 6.** Attenuation coefficient dependence on the frequency for fragments of canopies of different tree species: maple (M), aspen (A), and pine (P).

Fig. 4 shows that averaged attenuation coefficient dependencies on the frequency for the aspen and maple branches has a monotonic character and can be approximated by the following equation:

$$\mu = cf^{\beta} w_e \qquad (1)$$

where $\mu$ (dB/m) is the extinction coefficient, $w_e$ (kg/m$^3$) is the water content in the branches biomass. For the considered case, coefficients are equaled: $c=1.3$-$3.5$ and $\beta=0.6\pm0.2$ for aspen branches and $c=3.04$-$5.21$ and $\beta=0.55\pm0.17$ for maple branches. Value of $w_e$ is varied in the range of 1-3 kg/m$^3$ what is specific for natural conditions. There exist many retrieving algorithms for the vegetation water content based on microwave remote sensing observations, Normalized Difference Vegetation Index (NDVI), and Normalized Vegetation Water Index (NDWI). For example, the following approximation for the vegetation water content as function of the NDVI can be represented:

$$w_e H = \begin{cases} 1.9134(\text{NDVI})^2 - 0.3215(\text{NDVI}) & \text{when NDVI} \le 0.5; \\ 4.2857(\text{NDVI})^2 - 1.5429 & \text{when NDVI} > 0.5. \end{cases}$$

where $H$ is the canopy vertical size (m).

Curves of Fig. 5 show a dependence of attenuation coefficient both on the gravimetric vegetation moisture and on the type of vegetation fragments. In particular, attenuation coefficient increases from 0.2 dB/kg/m$^2$ to 2.0 dB/kg/m$^2$ monotonically when frequency changes in 0.8-4.0 GHz band and its value practically no changes in the 4.0-8.5 GHz band (curve 1). In the case of thick branches without leaves (curve 3), attenuation coefficient decreases with the frequency growth from 0.9 dB/kg/m$^2$ to 0.25 dB/kg/m$^2$. Thickness increase of the branches from 5 to 50 mm without leaves leads to the attenuation coefficient growth from 0.06 to 0.9 dB/kg/m$^2$ for low-frequency band, and to its decrease from 2.0 to 0.25 dB/kg/m$^2$ in high-frequency band. This effect is explained by that thin branches for long-waves represent media with efficient dielectric penetration closed to 1 but thickness branches are compared with wave-length what promotes to the resonance phenomena.

Results of Fig. 6 summarize the attenuation coefficient dependence on the frequency for different trees when their gravimetric moisture is changed from 42 to 52 percents. As result tree crones can attenuate electromagnetic waves with the following law: $\mu = 2.6 f^{0.44} w_e$. Fig. 7 represents dependencies of the attenuation coefficient on the gravimetric moisture of trees. From these results the following conclusion is followed that attenuation coefficient is directly proportional to the water content in the tree branches. Under this a slope of curves is increased with the frequency growth.

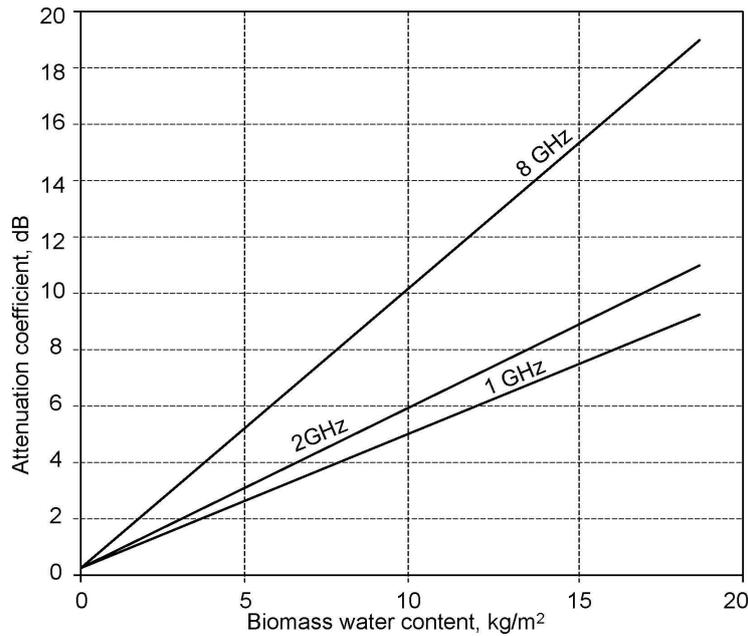
**Fig. 7.** Dependence of the attenuation coefficient on the canopy water content,

Finally, specific attenuation coefficient can be approximated by the formula (1) where coefficients $c$ and $\beta$ are defined by the vegetation type and canopy density. Principal contribution to the coefficients $c$ and $\beta$ is defined by the value of vegetation optical depth and other characteristics of the atmosphere-vegetation-soil system. Figure 8 represents dependence of $\beta$ on the gravimetric vegetation water content.

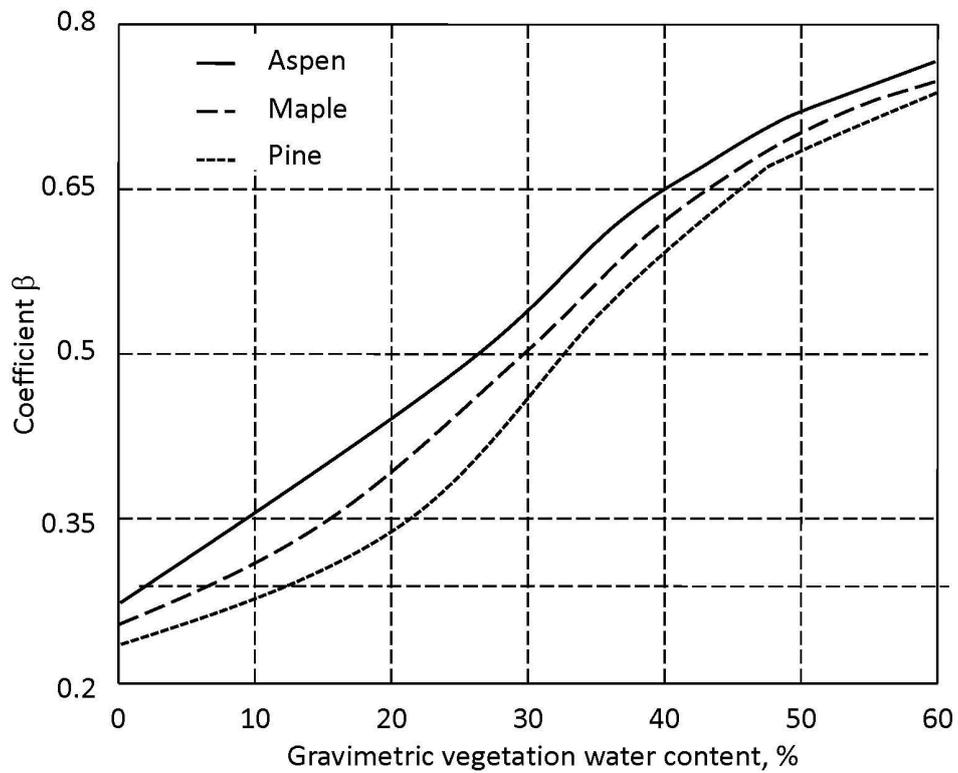
**Fig. 8.** A dependence of the coefficient $\beta$ on the gravimetric vegetation water content averaged by the branch diameters.

It should be noted that formula (1) described above illustrates a natural behaviour of variability[20-28]. Experimental data of several geophysical parameters have shown the same preference of variability depicting the nonlinear processes taking place during their variability. For example, atmospheric ozone and temperature exhibit non-linear behaviour [29-36]. This characteristic feature is used in several applications [37-39].

## 4. Theoretical generalization

The final purpose of this study is to introduce approach to estimation of the attenuation of microwaves under their propagation in the vegetation layer. Natural vegetation is characterized by the variety of types and canopy forms. The canopy of a tree consists of the mass of foliage and branches that are specified by spatial distribution and orientation, length, density, and categories as well as the water content and dielectric permittivity. Vegetation canopy structure has a significant impact on the microwave signal under its propagation in the vegetation layer. There exist different approaches to study this impact including 3D modeling tools coordinated with laboratory and in-situ observations [16]. During the MSRAMV measurements, typical canopies are formed in the camera what allows the evaluation of $c$ and $\beta$ coefficients represented in Table 2.

**Table 2.** Examples of attenuation models that summarize results of the measurements realized with the MSRAMV.

| Forest canopy type | The model (1) coefficients. | |
|---|---|---|
| | $c$ | $\beta$ |
| Young aspen forest with thin branches (0.5-1.0 cm)<br>    Duration of the vegetation period<br>    Winter time | 0.99±0.11<br>1.47±0.21 | 0.77±0.19<br>0.14±0.09 |
| Young maple forest with thin branches (0.5-1.0 cm)<br>    Duration of the vegetation period<br>    Winter time | 1.36±0.19<br>4.41±1.23 | 0.62±0.17<br>0.12±0.05 |
| Regular aspen forest during vegetation period: leaf radius is 1.9±0.5 cm, branches – radius is 4.5±0.8 cm, and length is 20.7±6.5 cm. | 1.13±0.22 | 0.48±0.18 |
| Regular maple forest during vegetation period: leaf radius is 5.1±0.7 cm, branches – radius is 3.9±0.6 cm, and length is 31.9±8.4 cm. | 1.78±0.45 | 0.43±0.16 |
| Pine forest: needle length is 7.2±0.4 cm, branches – radius is 2.8±0.7 cm, and length is 40.1±6.2 cm. | 0.61±0.18 | 0.35±0.12 |
| North-taiga forest: needle length is 2.9±0.3 cm, branches – radius is 3.2±0.6 cm, and length is 29.7±4.4 cm. | 1.32±0.19 | 0.44±0.14 |
| Traditional mixed forest of boreal zone. | 1.17±0.23 | 0.39±0.13 |

Coefficient $c$ in (1) is the function of the canopy parameters such as canopy volumetric density and dielectric constant $\varepsilon=\varepsilon_1-i\varepsilon_2$ that are specific for given tree type and $\varepsilon$ is function of temperature, salinity, frequency, vegetation bulk density and gravimetric vegetation water content. Real dielectric values can vary depending on the wood structure also. For example, according to [40] $\varepsilon_1$ and $\varepsilon_2$ of pine as function of the moisture content are changed in the ranges of 1.9-5.2 and 0.2-1.7, respectively.

Dielectric constant is changed with various depths in the tree body. Coefficient *β* determines a speed of the attenuation coefficient change depending on the frequency of electromagnetic wave [12]. Variety of such dependencies is defined by the diversity of canopies that is usually characterized by statistical distributions of foliages and branches. A model of efficient dielectric permittivity as function of the canopy density is given in [16]. For example, efficient dielectric permittivity for coniferous canopy is estimated by value of 2.008-$i$0.0006 for *f*=2.4 GHz. The main canopy parameter is optical depth $\tau$ that can by estimated with the use the following formula: $\tau=4pk_0\varepsilon_2 h/(3\cos\theta)$, where $k_0$ is the free space wave number, $h$ is linear dimension of the canopy in the direction $\theta$. In this case, total attenuation of microwaves under their propagation in the vegetation canopy will be equal to $b=\mu h$.

Model (1) allows the mapping attenuation effect for microwave propagation on given area covered by the vegetation of given type under the conditions when necessary parameters are estimation by existing monitoring system. Table 3 gives example of the attenuation of microwave at the frequency 1.2 GHz under $\theta=0°$ incidence in the forest canopy of some types. Seasonal parameters of these forests are estimated basing on the different publications Landsat 8 database with spatial resolution of 300 m and MODIS 250 m resolution.

**Table 3.** Seasonal distribution of averaged estimations of the canopy attenuation (dB) at the frequency *f*=1.2 GHz.

| Vegetation cover | Summer | Fall | Winter | Spring |
|---|---|---|---|---|
| North-taiga forest | 5.6 | 5.2 | 4.9 | 5.3 |
| Mid-taiga forest | 10.3 | 9.6 | 9.1 | 9.7 |
| South-taiga forest | 10.8 | 10.2 | 9.5 | 10.3 |
| Broad-leaved coniferous forest | 11.2 | 10.6 | 10.1 | 10.8 |
| Broad-leaved forest | 13.5 | 9.1 | 7.7 | 12.3 |
| Subtropical broad-leaved and coniferous forest | 12.9 | 12.7 | 12.5 | 12.7 |
| Humid evergreen tropical forest | 14.3 | 14.2 | 14.2 | 14.3 |
| Variable-humid deciduous tropical forest | 13.9 | 13.7 | 13.6 | 13.8 |

## 5. Conclusion and discussion

The paper provides new experimental tool for the study of microwave propagation through vegetation and model of 0.8-10 GHz narrowband radio signal attenuation in vegetation layer. The applicability of this attenuation model is based on knowledge and understanding of the propagation modes arising in vegetation layer and on a possibility to study experimentally these modes with the MSRAMV. The model was tested by comparing the modeling results and results of the measurements made with the MSRAMV and by radiative method in real situations [16]. Table 2 characterizes received precisions.

Undoubtedly, it is impossible during this work to cover all really probable situations of the microwaves propagation through vegetation layers. Really, the measuring system for retrieving attenuation of microwaves in vegetation (MSRAMV) was synthesized to be as experimental tool for the study of attenuation effects when radio waves of 0.8-10.0 GHz are propagated in the vegetation layer. The MSRAMV needs the knowledge of biometric vegetation characteristics such as canopy density and vegetation water content. Variety of these characteristics can be realized in the MSRAMV camera what allows the estimation of analytic attenuation model (1) two parameters. Application of this model does no need to form the canopy image by means of the representation of leaves and branches by discs and cylinders, respectively, what, certainly, connects with certain user mastery.

Of course, this user mastery is needed when the canopy image is formed within the MSRAMV camera but model (1) has only two unknown parameters as opposed to more complex models [9, 11, 18, 19, 41].

Model (1) can be improved at the expense of search of functional dependencies of its parameters on the vegetation biometric and physical characteristics such as biomass, geometry, density, structure, and optical depth that are estimated using existing monitoring systems [17, 42-44]. The results of this study show that simple model (1) together with MSRAMV measurements gives a possibility to assess the attenuation of microwaves under their propagation in the vegetation layer of various type in the arbitrary direction.